\documentclass[aps,prc,amsmath,amssymb,amsfonts,reprint,showpacs]{revtex4-1}
\usepackage{graphicx}
\begin{document}
\title{Saturation momentum scale extracted from semi-inclusive transverse spectra in high-energy pp collisions} 
\author{Takeshi Osada}
\email[]{t-osada@tcu.ac.jp}
\affiliation{
Department of Physics, 
Faculty of Liberal Arts and Sciences, \\ 
Tokyo City University, 
Tamazutsumi 1-28-1, Setagaya-ku, Tokyo 158-8557, Japan}
\author{Takuya Kumaoka}
\affiliation{Department of Physics, 
Shinshu University, 
Matsumoto 390-8621, Japan}
\date{\today}
\begin{abstract}
Geometric scaling is well confirmed for transverse momentum distributions observed in proton-proton collisions at LHC energies. 
We introduced multiplicity dependence on a saturation momentum of the geometrical scaling, 
assuming the scaling holds for semi-inclusive distributions as well as for inclusive distributions. 
The saturation momentum is usually given by Bjorken's $x$ variable, but redefinition of the scaling variable can make the saturation momentum 
a function of collision energy $W$. 
We treat the energy as a free parameter (denoted $W^*$ to distinguish it from $W$) and associate 
the energy-dependent saturation momentum $Q_{\rm sat}(W^*)$ with particle number density.   
By using $Q_{\rm sat}(W^*)$ for a scaling variable $\tau$, we show semi-inclusive distributions can be geometrically scaled. 
i.e., all semi-inclusive spectra observed at $W$=0.90, 2.76 and 7.00 TeV overlap one universal function.  
The particle density dependences of mean transverse momentum $\langle p_{\rm T} \rangle$ for LHC energies scales in terms of $Q_{\rm sat}(W^*)$. 
Furthermore, our model explains 
a scaling property of event-by-event $p_{\rm T}$ fluctuation measure $\sqrt{C_m}/\langle p_{\rm T}\rangle$ at LHC energies for pp collisions, where 
$C_m$ is two-particle transverse momentum correlator.  
Our analysis of the $p_{\rm T}$ fluctuation makes possible to evaluate a non-perturbative coefficient of the gluon correlation function. 
\end{abstract}
\maketitle
\section{Introduction}\label{sec:1}
Studies of small collision systems in high multiplicity events is attracting considerable interest \cite{Preghenella:2018moc} 
because of the collective phenomena which attribute to the formation of 
a strongly-interacting collectively-expanding quark-gluon medium \cite{Khachatryan:2010gv, Khachatryan:2016txc,Dusling:2015gta,Nagle:2018nvi}.  
A remarkable similarity has been observed between strange particles production in pp collisions and that in Pb-Pb collisions, 
suggesting the possibility of deconfined QCD phase formation in small systems \cite{ALICE:2017jyt}. 
In such pp collisions, the charged particle pseudo-rapidity density rises as a power of energy \cite{Kharzeev:2000ph,McLerran:2010ex}, 
which can be explained by the theory of gluon saturation \cite{Blaizot:1987nc, Gribov:1984tu}.  
Recombination of gluons \cite{Mueller:1985wy} in high particle number density state causes the saturation, 
and the gluon distribution function ceases growing from some intrinsic scale of the transverse momentum $Q_s$ \cite{Kharzeev:2001gp}. 
The Color Glass Condensate (CGC) \cite{Kovchegov:2012mbw,Iancu:2002xk,Blaizot:2004px} 
is an effective theory to describe saturated gluons with small $x$ as classical color fields 
radiated by color sources at higher rapidity. 
The existence of $Q_s$ which separates the degree of freedom into fast frozen color sources and slow dynamical color fields \cite{Gelis:2010nm} 
is the underlying assumption of the effective theory. 
The scaling of the limiting fragmentation curves \cite{Stasto:2011zza} 
is one of the crucial pieces of evidence for the picture of the CGC \cite{JalilianMarian:2002wq, McLerran:2004fg}. 

Another experimental evidence of CGC hypothesis is a geometrical scaling \cite{Iancu:2003jg,Iancu:2002tr} (GS) 
confirmed originally in results on total $\gamma^*$p cross section \cite{Stasto:2000er}. 
A term of the `geometrical' of this GS comes from that survival probability of a color dipole \cite{Tribedy:2010ab,Iancu:2003ge,Gotsman:2019vrv}
is determined by the geometric relationship between the dipole size and the saturation radius given by $Q_{s}^{-1}(x)$ \cite{Gelis:2010nm,Munier:2003vc}, 
where $x$ is a Bjorken variable. 
In this article, since we will deal with multi-particle production in the central rapidity region of high-energy pp collisions, 
we have $x=p_{\rm T}/W$, where $p_{\rm T}$ and $W$ are transverse momentum and colliding energy of the incident proton, respectively. 
With $x_0$, $Q_0$ and $\lambda$ are constants (see, Sec.\ref{sec:2} for details), the saturation momentum is given by \cite{GolecBiernat:1998js}  
\begin{eqnarray}
Q_{s}(x) \equiv Q_0 \left( \frac{x}{x_0} \right)^{-\lambda/2}. 
\end{eqnarray}
If such momentum is the only scale that controls $p_{\rm T}$ distribution, it should exhibit GS behavior; i.e., 
when one normalizes inclusive transverse momentum spectra observed with an appropriate constant $S_{\rm T}$ (interpreted later 
as reaction effective transverse cross-sectional area), 
the data points lies on a characteristic curve ${\cal F}(\tau)$ which is only depends on the scaling variable $\tau\equiv p^2_{\rm T}/Q^2_{\rm sat}$ 
and the curve does not depend on $W$. 
In particular, the scaling property has been vigorously studied for 
pp collisions obtained at the Large Hadron Collider (LHC) energies \cite{Praszalowicz:2015dta, Praszalowicz:2013fsa, Praszalowicz:2011tc,Praszalowicz:2011rm}  
and GS is observed in single inclusive distributions of charged hadrons \cite{Praszalowicz:2015dta} and 
and recently observed direct photons from heavy-ion collisions \cite{Khachatryan:2019uqn}.  
Since $Q_s(x)$ includes $p_{\rm T}$ dependence via Bjorken $x$; i.e., 
\begin{eqnarray}
Q_s(\frac{p_{\rm T}}{W})= Q_0\cdot (x_0 W)^{\frac{\lambda}{2}} 
~p_{\rm T}^{-\frac{\lambda}{2}}, 
\end{eqnarray}
we unify the terms of $p_{\rm T}$ contained in $\tau$ 
and redefine the rest that depends on $W$ as an energy dependent saturation momentum $Q_{\rm sat}(W)$ \cite{Kharzeev:2004if}. 
Namely, the scaling variable can be rewritten as  
\begin{subequations}
\begin{eqnarray}
   && \tau = \left( \frac{p_{\rm T}}{Q_{\rm sat}(W)} \right)^{2+\lambda}, \label{eq:tau_pt_Qsat} \\ 
   && Q_{\rm sat}(W) \equiv Q_0 \left( \frac{x_0 W}{Q_0}\right)^{\frac{\lambda}{\lambda+2}} \label{eq:Qsat_def}
\end{eqnarray} 
and the GS is expressed as \cite{McLerran:2014apa}
\begin{eqnarray}
   \frac{1}{S_{\rm T}}~\frac{d^2N_{\rm ch}}{dp^2_{\rm T} dy}
     = {\cal F}\left( \tau \right), 
   \label{eq:GS_inclusive}
\end{eqnarray}
where ${\cal F}(\tau)$ is a so-called universal function of GS. 
\end{subequations}
Under an assumption that a local parton hadron duality \cite{Azimov:1984np} as a hadronization model is appropriate,  
the particle density at the central rapidity region ($y\approx 0$) relates to $Q_{\rm sat}(W)$ as follows:  
\begin{eqnarray}
  \left\langle \frac{dN_{\rm ch}}{dy} \right\rangle 
  \propto ~S_{\rm T}~Q_{\rm sat}^2(W), \label{eq:Multiplicity_averageSaturationScale}
\end{eqnarray} 
where $\langle \cdots \rangle$ denotes 
the average over single inclusive distribution (or over minimum bias events). 
Since the particle number density is known to increases gradually with collision energy $W$, we expect $Q_{\rm sat}(W)$ to also increases gradually with $W$. 

Let us suppose that GS holds not only for inclusive distributions but also for the {\it semi-inclusive} distributions, 
i.e., inclusive distribution with fixed multiplicity or limited multiplicity class \cite{Kanki:1988tz} 
\footnote{In this paper, the inclusive spectra is denoted by $d^2N_{\rm ch}/dp_{\rm T} dy$, 
and the semi-inclusive spectra is denoted by $d^2n_{\rm ch}/dp_{\rm T} dy$ to distinguish it from the inclusive one.}. 
For the semi-inclusive spectra $\frac{d^2n_{\rm ch}}{dp_{\rm T}^2 dy}$, as the case of inclusive one, we assume that there exists a saturation momentum 
for the spectrum classified by multiplicity as well and we propose to represent it by effective energy $W^*$; i.e., 
\begin{subequations}
\begin{eqnarray}
    \frac{1}{S_{\rm T}^*}~\frac{d^2n_{\rm ch}}{dp_{\rm T}^2 dy} = {\cal F}\left(\tau\right), 
    \label{eq:Tsallis-n_fix}
\end{eqnarray}
where, instead of Eq.(\ref{eq:tau_pt_Qsat}), we use 
\begin{eqnarray}
 \tau = \left( \frac{p_{\rm T}}{Q_{\rm sat}(W^*)} \right)^{2+\lambda}. 
   \label{eq:W**}
\end{eqnarray} 
\end{subequations}
It should be noted here that the universal function ${\cal F}$ in Eq.(\ref{eq:Tsallis-n_fix}) is the same as that in Eq.(\ref{eq:GS_inclusive}). 
Here, $S^*_{\rm T}$ and $W^*$ are determined to reproduce the spectrum obtained by the experiment.
In particular, this $W^*$ is a fit parameter introduced replacing the actual beam energy $W$ in eq.(\ref{eq:Qsat_def}). 
Hence, we intend to check whether GS found in inclusive distribution is restored even in semi-inclusive distribution. 

It may be appropriate to give some explanations for $W^*$ here. 
As discussed in detail later in Sec.\ref{sec:2}, the energy-dependent saturation momentum $Q_{\rm sat}$ gives a typical scale 
of transverse momentum $p_{\rm T}$. That is, $Q_{\rm sat}$ is the solution $p_{\rm T}$ of an equation $Q_s(p_{\rm T}/W)=p_{\rm T}$ 
for each colliding energy $W$. 
Because $Q_{\rm sat}$ itself is a scale of transverse momentum, 
the inverse of it is a typical transverse size scale of saturated gluons.  
Hence as seen in Eq.(\ref{eq:Multiplicity_averageSaturationScale}), 
the ratio of effective interaction cross sectional area $S_{\rm T}$ to the cross-sectional area per gluon $Q_{\rm sat}^{-2}$ 
governs the mean charged particle density of the inclusive distribution.   
On the other hand, for semi-inclusive collisions classified by multiplicity, 
$S^*_{\rm T}$  and $Q_{\rm sat}(W^*)$ should be related to each other by the constraint of the fixed multiplicity. 
We will discuss the relation between $Q_{\rm sat}(W^*)$ and $S_{\rm T}^*$ of the semi-inclusive distribution 
in some detail in Sec.\ref{sec:3} and also comment on the physical meaning of $W^*$.  

This article is organized as follows. 
In the following Sec.\ref{sec:2}, we briefly review GS hypothesis and we confirm that it holds well 
for inclusive transverse spectra observed in pp collisions at LHC energies. Then, we determine the 
universal function of GS used throughout this article. 
In Sec.\ref{sec:3}, the effective energy $W^*$ is determined from the semi-inclusive transverse spectra. 
By using the scaling variables with $Q_{\rm sat}(W^*)$, we show that the transverse momentum spectra observed in the different multiplicity classes at the different collision energies scale to the 
universal function found in Sec.\ref{sec:2}. 
We also show that the multiplicity dependence of the mean transverse momentum scales with $Q_{\rm sat}(W^*)$. 
Furthermore, we analyze the scaling behavior of a normalized fluctuation measure of transverse momentum and 
consider it as a result of the correlation between particles generated from color flux tubes.  
We close with Sec.\ref{sec:4} containing the summary and some concluding remarks. 
\section{GS for inclusive $p_{\rm T}$ distribution}\label{sec:2}
\begin{figure}
\centerline{\includegraphics[width=10.0cm]{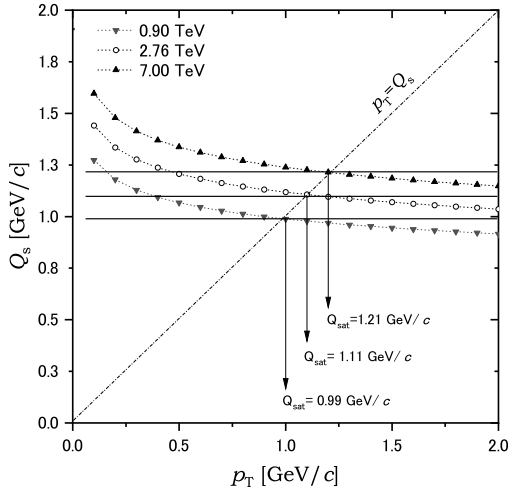}}
\caption{Saturation momentum $Q_s$ (dotted curve with triangle and circle symbols) 
and energy-dependent saturation momentum $Q_{\rm sat}$ (horizontal solid lines) for $W=$ 0.90, 2.76 and 7.00 TeV. The long and short dashed line represents $Q_s=p_{\rm T}$. 
The intersections of the line and the dotted curve give $Q_{\rm sat}$ at each $W$.  
For $W=0.90, 2.76$ and $7.00$ TeV, $Q_{\rm sat}=0.99, ~1.11, ~1.21$~GeV/$c$, respectively.
}\label{fig:1}
\end{figure}

\begin{figure}
\centerline{\includegraphics[width=10.0cm]{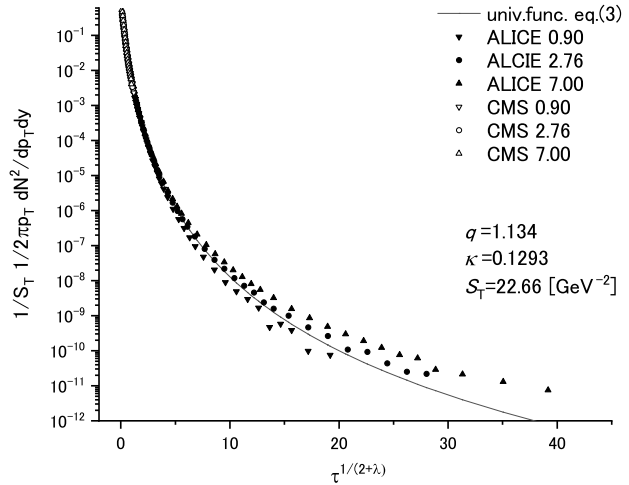}}
\caption{
The transverse momentum distributions exhibit geometrical scaling behavior for pp collisions at $W$=0.90, 2.76 and 7.00 TeV.  
Experimental data (indicated by triangles or circles) are observed by 
ALICE Collaboration\cite{Abelev:2013ala} and CMS Collaboration\cite{Chatrchyan:2012qb}.
The solid curve is the universal function ${\cal F}(\tau)$ with $q=$1.134,  $\kappa=$0.1293 and $\lambda=$0.22 (See Eq.(\ref{eq:universal_func_fit})).  
The effective interaction cross-sectional area $S_{\rm T}=$22.66~GeV$^{-2}$ is used.}\label{Fig:2}
\end{figure}

The transverse momentum spectra of various energies for pp collisions never scale with variable $p_{\rm T}$ 
because their intensities and slopes depend on the colliding energy $W$. However, for high energy collisions 
in which the number of soft gluons inside the proton saturates, the transverse momentum spectrum depends only 
on a scaling variable defined by Eq.(\ref{eq:tau_pt_Qsat}) with Eq.(\ref{eq:Qsat_def}).  
Let us examine the quantitative difference between $Q_s(x)$ and $Q_{\rm sat}(W)$ at LHC energies.  
We show them as a function of $p_{\rm T}$ in Fig.\ref{fig:1} for the case of 
$\lambda=0.22$, $x_0=1.0\times10^{-3}$, $Q_0=1.0$ GeV/$c$ \cite{McLerran:2014apa}, 
and we will fix the values from now on. 
Since $Q_s$ is less dependent on $p_{\rm T}$ for $p_{\rm T}\gtrsim 0.5$ GeV/$c$, 
one may use $Q_{\rm sat}(W)$ instead of $Q_s$ as a typical momentum scale.  
The values of $Q_{\rm sat}(W)$ obtained from inclusive $p_{\rm T}$ spectra 
at energy $W=$ 0.90, 2.76, 7.00 TeV are 0.99, 1.11 and 1.21 GeV/$c$, respectively. 
As shown in  Fig.\ref{Fig:2}, experimental data observed by ALICE \cite{Abelev:2013ala} and CMS Collaboration \cite{Chatrchyan:2012qb} 
suggests the validity of GS especially for $\tau^{1/(2+\lambda)} \lesssim 10$.  
The curve emerging from a plot of the $p_{\rm T}$ spectra with using the scaling variable $\tau$ 
can be fitted well by the so-called Tsallis type function \cite{Rybczynski:2012pn,Rybczynski:2012vj}; 
\begin{eqnarray}
  {\cal F}(\tau)= \left[1+(q-1)\frac{\tau^{1/(\lambda+2)}}{\kappa} \right]^{-1/(q-1)},
\label{eq:universal_func_fit} 
\end{eqnarray}
where the non-extensive parameter $q$=1.134 and  $\kappa$=0.1293 are used. 
The effective cross-sectional area in Eq.(\ref{eq:GS_inclusive}) is determined as $S_{\rm T}=$22.66 GeV${}^{-2}$.  
In this way, the transverse momentum distribution indeed exhibits GS behavior for pp collisions in the LHC energies. 
It seems to be appropriate to shortly comment on the energy-dependent saturation momentum $Q_{\rm sat}$ 
and an effective temperature $T_{\rm eff}$ (or a slope parameter) \cite{Praszalowicz:2013fsa,McLerran:2013una} here. 
In case of Tsallis-type distribution function, $T_{\rm eff}$ can be defined as 
\begin{eqnarray}
    \frac{1}{2\pi p_{\rm T}} \frac{d^2N_{\rm ch}}{dp_{\rm T} dy}
    =C \left[1+(q-1)\frac{p_T}{T_{\rm eff}} \right]^{-1/(q-1)}. \label{eq:Tsallis-type}
\end{eqnarray}
Here, one may interpret the constant $C$ as $S_{\rm T}$. 
Since the transverse spectra experimentally observed exhibits good GS behavior,  
the effective temperature $T_{\rm eff}$ 
must have energy dependence to cancel 
the energy dependence of  
\begin{eqnarray}
  p_{\rm T} = Q_0 \left(\frac{x_0W}{Q_0}\right)^{\frac{\lambda}{\lambda+2}} \tau^{\frac{1}{2+\lambda}}, 
  \label{eq:pt_expression}
\end{eqnarray}
which is obtained from Eq.(\ref{eq:tau_pt_Qsat}). 
Hence, the property of the GS determined the energy dependence of $T_{\rm eff}$ and 
that $T_{\rm eff}$ should be proportional to $Q_{\rm sat}$ \cite{Praszalowicz:2013fsa}.
Substituting Eq.(\ref{eq:pt_expression}) into Eq.(\ref{eq:Tsallis-type}) yields the expression of the 
universal function of Eq.(\ref{eq:universal_func_fit}) in the case of 
\begin{eqnarray}
T_{\rm eff}=\kappa Q_{\rm sat}.\label{eq:T=kQ}
\end{eqnarray}  
The gluon saturation is physics of the intermediate energy scale $Q_{\rm sat}$, while GS observed in the final state is physics of energy scale $T_{\rm eff}$ which is 
much lower than $Q_{\rm sat}$. Therefore, the parameter $\kappa$ in Eq.(\ref{eq:T=kQ}) (or equivalently Eq.(\ref{eq:universal_func_fit})) 
may have a physical meaning of a linkage between two energy scales of $Q_{\rm sat}$ and $T_{\rm eff}$. 
Before closing Sec.\ref{sec:2}, let us check $S_{\rm T}$ and $Q_{\rm sat}$ obtained here. 
By integrating Eq.(\ref{eq:Tsallis-type}), we obtain the average multiplicity density\cite{Osada:2017oxe}; 
\begin{eqnarray}
  \left\langle \frac{dN_{\rm ch}}{dy} \right\rangle &=&  \frac{2\pi S_{\rm T}[\kappa Q_{\rm sat}]^2}{(2-q)(3-q)} = \frac{3.76}{[{\rm GeV/}^{2}]}~Q_{\rm sat}^2,     
\end{eqnarray}
which gives 3.68, 4.63, and 5.50 for $W=$0.90, 2.76 and 7.00 TeV, respectively. 
These values should be compared with values obtained by experiments \cite{Adam:2015gka} i.e., 
3.75${}^{+0.06}_{-0.05}$, 4.76${}^{+0.08}_{-0.07}$, 5.98${}^{+0.09}_{-0.07}$ for $\sqrt{s}=$ 0.90, 2.76 and 7.00 TeV, respectively. 
\section{GS for semi-inclusive $p_{\rm T}$ distribution}\label{sec:3}
\subsection{Extraction of saturation momentum scale}\label{sec:3.1}
In this Section, we will extract the multiplicity dependence of saturation momentum $Q_{\rm sat}$ from the semi-inclusive spectrum observed. 
Our central assumption is that the semi-inclusive $p_{\rm T}$ distribution scales to the same universal function ${\cal F}(\tau)$ 
as the inclusive one (i.e., Eq.(\ref{eq:universal_func_fit}) with $q=1.134$,  $\kappa=$0.1293 and $\lambda=$0.22), 
providing that the appropriate $Q_{\rm sat}(W^*)$ is used. 
Since $S_{\rm T}$ in Eq.(\ref{eq:GS_inclusive}) now depends on the multiplicity, we require GS for 
the semi-inclusive spectra as shown by Eq.(\ref{eq:Tsallis-n_fix}) with (\ref{eq:W**}) in Sec.\ref{sec:1}; 
\begin{eqnarray*}
    \frac{1}{S^*_{\rm T}}~\frac{1}{2\pi p_{\rm T}} \frac{d^2n_{\rm ch}}{dp_{\rm T} dy}
    ={\cal F}(\tau), 
\end{eqnarray*}
and 
\begin{eqnarray*}
   \tau^{1/(2+\lambda)}=\frac{p_{\rm T}}{Q_{\rm sat}(W^*)}.
\end{eqnarray*} 
These two parameters, $W^*$ and $S^*_{\rm T}$, are determined by fitting to the experimental data on the semi-inclusive $p_{\rm T}$ distribution. 
Note that, in this case, Eq.(\ref{eq:Multiplicity_averageSaturationScale}) should be modified as 
\begin{eqnarray}
  \frac{dn_{\rm ch}}{dy} 
  \propto ~S^*_{\rm T}~Q_{\rm sat}^2(W^*). \label{eq:dndy_semi-inclusive}
\end{eqnarray}
Since the universal function in Eq.(\ref{eq:Tsallis-n_fix})  is the same as that in Eq.(\ref{eq:GS_inclusive}),  
the proportionality constants in Eqs.(\ref{eq:Multiplicity_averageSaturationScale}) and (\ref{eq:dndy_semi-inclusive}) are equal. 
Therefore, using Eq.(\ref{eq:Qsat_def}), the ratio of $W^*$ to $W$ is given by 
\begin{eqnarray}
  \frac{W^*}{W} &=& \left[ 
  \frac{S_{\rm T}/ \langle \frac{dN_{\rm ch}}{dy}\rangle}
         {S^*_{\rm T}/ \frac{dn_{\rm ch}}{dy}} 
  \right]^{\frac{2+\lambda}{2\lambda}}\nonumber \\ 
 &=&\left\{ 
\begin{array}{rl}
 \Big[0.23~[{\rm fm}^2]/s^*_{\rm T} \Big]^{5.05}&  (W=0.90~{\rm TeV}), \\
 \Big[0.18~[{\rm fm}^2]/s^*_{\rm T} \Big]^{5.05}&  (W=2.76~{\rm TeV}), \\
 \Big[0.16~[{\rm fm}^2]/s^*_{\rm T} \Big]^{5.05}&  (W=7.00~{\rm TeV)}, 
 \end{array}
 \right. \label{eq:W*perW_dndy} 
\end{eqnarray}
where $s^*_{\rm T}\equiv S^*_{\rm T}/ \frac{dn_{\rm ch}}{dy}$.
As can be seen from Eq.(\ref{eq:W*perW_dndy}), $W^*$ and $S^*_{\rm T}$ are not independent parameters.
Hence, whether $W^*/W$ becomes larger or smaller than unity depends on 
whether a cross-sectional area per gluon $s^*_{\rm T}$ in the semi-inclusive distribution 
is larger or smaller than that in the inclusive distribution. Even if $W^*$ has a value greater than $W$, it does not mean 
an unphysical situation. 

\begin{figure}
\centerline{\includegraphics[width=9.0cm]{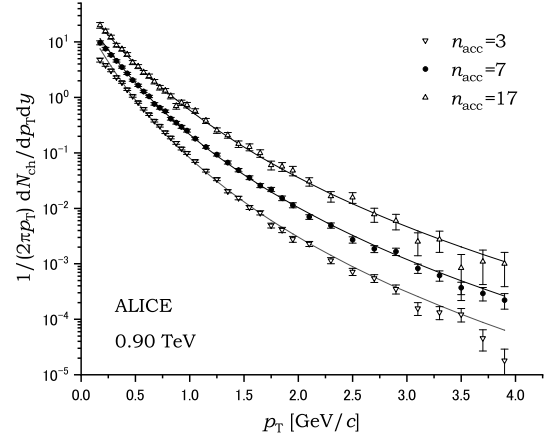}}
\caption{Fit results of $S^*_{\rm T}{\cal F}$ (solid curves; see Eqs.(\ref{eq:Tsallis-n_fix}) and (\ref{eq:W**})) to 
$p_{\rm T}$ spectra with $n_{\rm acc}$=3, 7 and 17 in pp collisions at energy 0.90 TeV observed by ALICE Collaboration \cite{Aamodt:2010my}. 
The pseudo-rapidity range is $-0.8\le \eta \le +0.8$.}\label{fig:3}
\end{figure}

\begin{figure}
\centerline{\includegraphics[width=10.0cm]{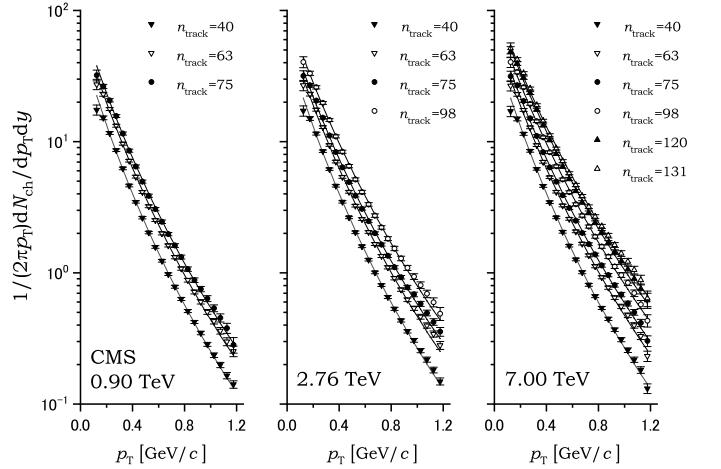}}
\caption{
The same as Fig.\ref{fig:3} but to data on 
$p_{\rm T}$ spectra for multiplicity selections with $n_{\rm tracks}=40\sim131$ in pp collisions at energy 
0.90, 2.76 and 7.00 TeV observed by CMS Collaboration \cite{Chatrchyan:2012qb}. 
The pseudorapidity range is $-2.4\le \eta \le +2.4$.  
}\label{fig:4}
\end{figure}

\begin{figure*}
\centerline{\includegraphics[width=13.0cm]{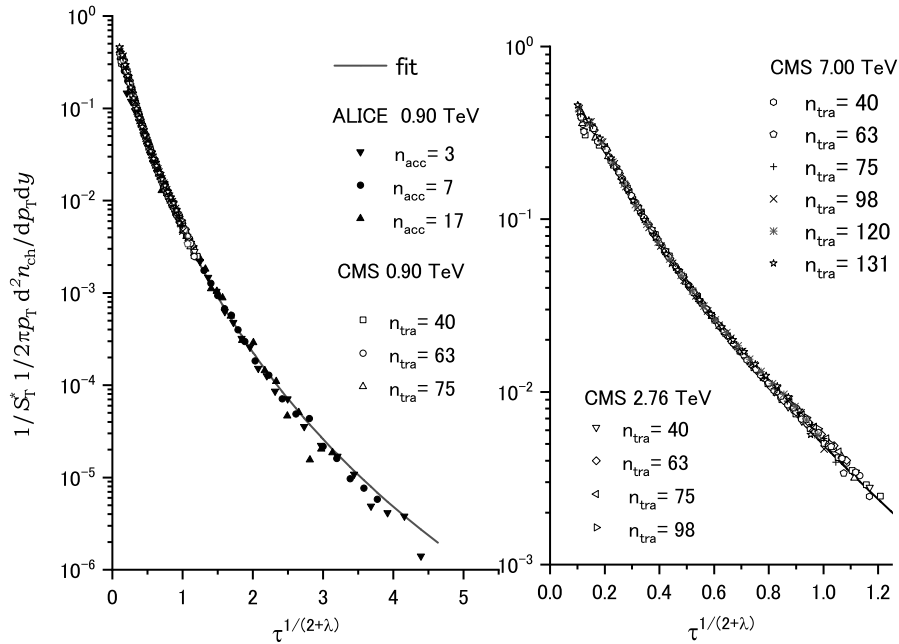}}
\caption{Geometrical scaling of the semi-inclusive transverse momentum spectra 
in terms of the scaling variable $\tau$ defined by Eq.(\ref{eq:W**}). 
The experimental data are observed by ALICE \cite{Aamodt:2010my} (open symbols) with multiplicity class $n_{\rm acc}$ = 3, 7, 17 and by 
CMS \cite{Chatrchyan:2012qb} (closed symbols) with multiplicity class $n_{\rm tracks}$ = 40, 63, 75, 98, 120 and 131. 
}\label{fig:5} 
\end{figure*}

\begin{figure}
\centerline{\includegraphics[width=9.0cm]{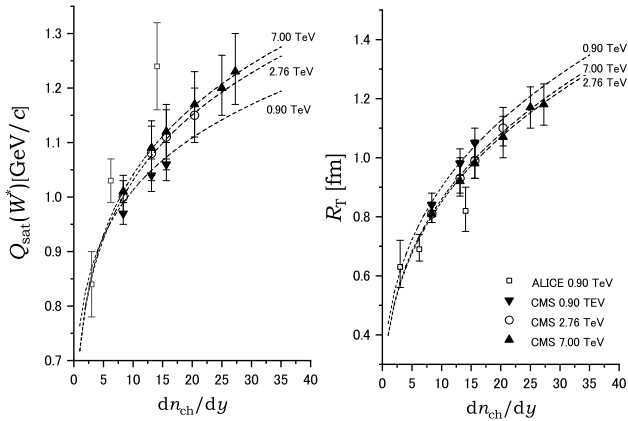}}
\caption{
The energy-dependent saturation momentum $Q_{\rm sat}(W^*)$ (left panel) and effective interaction radius $R_{\rm T}$ (right panel) 
extracted from the semi-inclusive transverse spectra for pp collisions at energy 0.90, 2.76 and 7.00 TeV.  
The range of $\chi^2<1.5\chi^2_{\rm min}$ is represented as error bars. 
The dashed curves are fit results. (See Eq.(\ref{eq:Qsat_dndy}) and (\ref{eq:Rt_dndy}).)}\label{fig:6}
\end{figure}

In order to determine the multiplicity dependence of $W^*$ in Eq.(\ref{eq:W**}), 
we fit Eq.(\ref{eq:Tsallis-n_fix}) to $p_{\rm T}$ spectra at energy 0.90 TeV for 
the accepted number of charged particles $n_{\rm acc}=$ 3, 7 and 17 observed ALICE Collaboration \cite{Aamodt:2010my}  
and at energy 0.90, 2.76 and 7.00 TeV for the average track multiplicity $n_{\rm tracks}=$ 40, 63, 75, 98, 120 and 131 observed CMS Collaboration \cite{Chatrchyan:2012qb}. 
Figure \ref{fig:3} and \ref{fig:4} show the results of fitting with $S^*_{\rm T}{\cal F}$ to ALICE and CMS data, respectively.
Besides, Table \ref{tab:w-ast} shows the values of  $W^{*}$ (multiplied by $x_0$) and effective radius $R_{\rm T}\equiv \sqrt{S^*_{\rm T}/\pi}$ obtained by the fit. 
Table \ref{tab:w-ast} also shows the value of $Q_{\rm sat}(W^*)$ and the minimum value of $\chi^2$ (denoting by $\chi^2_{\rm min}$) in each fitting.
\begin{table*}
\caption{\label{tab:w-ast} The values of $x^0W^*$, $Q_{\rm sat}$, $R_{\rm T}$, and minimum chi-squared $\chi^2_{\rm min}$ 
obtained from the fitting to the semi-inclusive transverse momentum distribution observed by 
ALICE \cite{Aamodt:2010my} with multiplicity class $n_{\rm acc}$ = 3, 7, 17 (accepted number of charged particles per inelastic event in the range $|\eta|<0.8$) 
and by CMS \cite{Chatrchyan:2012qb} with multiplicity class $n_{\rm tracks}$ = 40, 63, 75, 98, 120 and 131(average number of true tracks multiplicity in the range $|\eta|<2.4$). 
For assignment from $n_{\rm acc}$ to $\langle n_{\rm ch} \rangle$ in ALICE data, we use results presented in Table 2 of Ref.\cite{Aamodt:2010my}.  
The particle densities $dn_{\rm ch}/dy$ at central rapidity region, are estimated by $\langle n_{\rm ch}\rangle/\Delta\eta$ for simplicity.  
${X}^{+a}_{-b}$ denotes $X (=x_0W^*, Q_{\rm sat}, R_{\rm T})$ giving $\chi^2_{\rm min}$ and $a$, $b$ mean a boundary $X-b \leqq X \leqq X+a$ giving $\chi^2 = 1.5~\chi^2_{\rm min}$. } 
\begin{ruledtabular}
\begin{tabular}{crcccc r}
$~\sqrt{s}~$(TeV)&$n_{\rm acc}$&$\langle n_{\rm ch}\rangle/\Delta\eta$&  
~$x_0W^*$ [GeV]  &
~$Q_{\rm sat}$[GeV/$c$] &
~$R_{\rm T}$ [fm]  & 
~$\chi^2_{\rm min}$/dof  \\ 
\hline
0.90 &3 & 4.8/1.6  & 0.18$^{+0.16}_{-0.09}$  & 0.84$^{+0.06}_{-0.06}$  & 0.63$^{+0.09}_{-0.07}$ & 97.5/33 \\     
0.90 &7 &10.0/1.6 & 1.39$^{+0.60}_{-0.43}$  & 1.03$^{+0.04}_{-0.04}$  & 0.69$^{+0.05}_{-0.04}$ & 20.5/33 \\      
0.90&17 &22.5/1.6 & 9.03$^{+7.70}_{-4.32}$ & 1.24$^{+0.08}_{-0.08}$  & 0.82$^{+0.08}_{-0.07}$ & 23.2/33 \\     
$~\sqrt{s}~$(TeV)&$n_{\rm tracks}$& $\langle n_{\rm ch}\rangle/\Delta\eta$& 
$x_0W^*$ [GeV]  &
~$Q_{\rm sat}$[GeV/$c$] &
$R_{\rm T}$ [fm]  & 
$\chi^2_{\rm min}$/dof  \\ 
\hline 
0.90&40 & $40/4.8$   & 0.77$^{+0.16}_{-0.14}$ & 0.97$^{+0.02}_{-0.02}$ & 0.84$^{+0.04}_{-0.02}$  & 10.5/18 \\ 
0.90&63 & $63/4.8$   & 1.45$^{+0.53}_{-0.39}$ & 1.04$^{+0.03}_{-0.03}$ & 0.98$^{+0.05}_{-0.04}$  & 21.8/18 \\ 
0.90&75 & $75/4.8$   & 1.72$^{+0.22}_{-0.45}$ & 1.06$^{+0.04}_{-0.03}$ & 1.05$^{+0.05}_{-0.05}$  & 17.6/18 \\ \hline 
2.76&40 & $40/4.8$  & 1.05$^{+0.33}_{-0.24}$ & 1.00$^{+0.03}_{-0.03}$ & 0.81$^{+0.04}_{-0.03}$  & 19.8/18 \\ 
2.76&63 & $63/4.8$  & 2.23$^{+1.15}_{-0.80}$ & 1.08$^{+0.05}_{-0.05}$ & 0.93$^{+0.07}_{-0.05}$  & 41.0/18 \\ 
2.76&75 & $75/4.8$  & 2.92$^{+1.79}_{-1.17}$ & 1.11$^{+0.05}_{-0.06}$ & 0.99$^{+0.07}_{-0.06}$  & 46.4/18 \\ 
2.76&98 & $98/4.8$  & 3.94$^{+2.24}_{-1.54}$ & 1.15$^{+0.05}_{-0.05}$ & 1.10$^{+0.07}_{-0.06}$  & 32.4/18 \\ \hline 
7.00&40 & $ 40/4.8$   & 1.06$^{+0.35}_{-0.19}$ & 1.01$^{+0.03}_{-0.02}$ & 0.81$^{+0.03}_{-0.03}$  & 19.1/18 \\ 
7.00&63 & $ 63/4.8$   & 2.47$^{+1.25}_{-0.88}$ & 1.09$^{+0.05}_{-0.05}$ & 0.92$^{+0.07}_{-0.05}$  & 41.1/18 \\ 
7.00&75 & $ 75/4.8$   & 3.24$^{+1.71}_{-1.17}$ & 1.12$^{+0.05}_{-0.05}$ & 0.98$^{+0.07}_{-0.05}$  & 40.9/18 \\ 
7.00&98 & $ 98/4.8$   & 4.95$^{+3.20}_{-2.00}$ & 1.17$^{+0.06}_{-0.06}$  & 1.07$^{+0.07}_{-0.07}$ & 47.1/18 \\ 
7.00&120 & $120/4.8$& 6.25$^{+3.92}_{-2.17}$ & 1.20$^{+0.06}_{-0.05}$ & 1.17$^{+0.07}_{-0.07}$  & 30.1/18 \\ 
7.00&131 & $131/4.8$& 8.25$^{+5.67}_{-3.25}$ & 1.23$^{+0.07}_{-0.06}$ & 1.18$^{+0.07}_{-0.07}$  & 35.6/18 \\ 
\end{tabular}
\end{ruledtabular}
\end{table*}
As shown in Fig.\ref{fig:5}, we confirm that the semi-inclusive transverse momentum spectra depicted in Figs.\ref{fig:3} and \ref{fig:4} 
scale in terms of the scaling variable $\tau$ of Eq.(\ref{eq:W**}).
Note that the solid curve (the universal function ${\cal F}$) in Fig.\ref{fig:5} 
is exactly the same as that obtained in the inclusive distribution in Fig.\ref{fig:1}. 
We also show $Q_{\rm sat}$ and $R_{\rm T}$ as function of $dn_{\rm ch}/dy$ in Fig.\ref{fig:6}. 
It is found that $Q_{\rm sat}(W^*)$ and $R_{\rm T}$ are proportional to $\left(dn_{\rm ch}/dy\right)^{1/6}$ and $\left(dn_{\rm ch}/dy\right)^{1/3}$, respectively.
The curves depicted by broken lines in the left panel (for $Q_{\rm sat}$) and the right panel (for $R_{\rm T}$) of Fig.\ref{fig:6} are given by 
\begin{subequations}
\begin{eqnarray}
\frac{Q_{\rm sat}}{[{\rm GeV}/c]}&=&\left\{ 
\begin{array}{rl}
0.232 +0.532 \left( \frac{dn_{\rm ch}}{dy}\right)^{\frac{1}{6}} &  (W=0.90~{\rm TeV}), \\
0.049 +0.669 \left( \frac{dn_{\rm ch}}{dy}\right)^{\frac{1}{6}} &  (W=2.76~{\rm TeV}), \\
0.031 +0.688 \left( \frac{dn_{\rm ch}}{dy}\right)^{\frac{1}{6}} &  (W=7.00~{\rm TeV)}, \nonumber
\end{array}
\right. \label{eq:Qsat_dndy}\\ 
\end{eqnarray}
\begin{eqnarray}
\frac{R_{\rm T}}{[{\rm fm}]}&=&\left\{ 
\begin{array}{rl}
0.039 +0.400 \left( \frac{dn_{\rm ch}}{dy}\right)^{\frac{1}{3}} & (W=0.90~{\rm TeV}), \\
0.006 +0.396 \left( \frac{dn_{\rm ch}}{dy}\right)^{\frac{1}{3}} & (W=2.76~{\rm TeV}), \\
0.006 +0.392 \left( \frac{dn_{\rm ch}}{dy}\right)^{\frac{1}{3}} & (W=7.00~{\rm TeV}).\nonumber 
\end{array}
\right.  \label{eq:Rt_dndy}\\ 
\end{eqnarray}
\end{subequations}
These $dn_{\rm ch}/dy$ dependencies are consistent with Eq.(\ref{eq:dndy_semi-inclusive}) 
when $dn_{\rm ch}/dy$ is sufficiently large and the constant term can be ignored.
Here, it is interesting to find a particle number density $dn_{\rm ch}/dy$ to give $W^*(dn_{\rm ch}/dy)=W$. 
Using Eqs.(\ref{eq:W*perW_dndy}) and (\ref{eq:Rt_dndy}), we can evaluate $s^*_{\rm T}$ that satisfies $W^*/W=1$. 
For simplicity, ignoring the constant term of Eq.(\ref{eq:Rt_dndy}), we obtain $W^*=W$ when 
$dn_{\rm  ch}/dy=$10.4, 20.5 and 27.4 for 0.90, 2.76 and 7.00 TeV, respectively. 
In fact, for CMS event classes with $n_{\rm tracks}$ = 63, 98 and 131 in $|\Delta \eta|<2.4$ 
at $W=$0.90, 2.76 and 7.00 TeV, respectively, it can be read from Table \ref{tab:w-ast} that $W^*>W$ is realized. 
\subsection{Mean transverse momentum}
\begin{figure}
\centerline{\includegraphics[width=9.0cm]{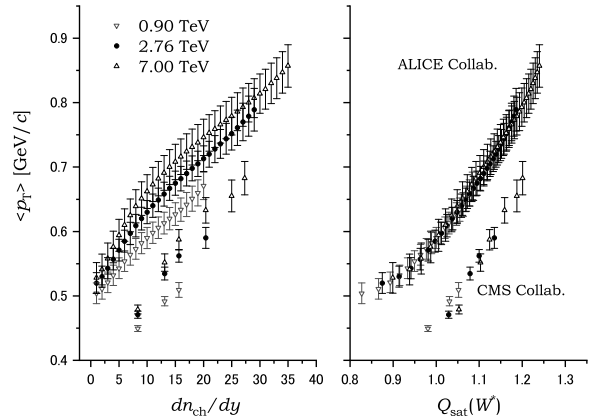}}
\caption{Average transverse momentum $\langle p_{\rm T} \rangle$ of pp collisions observed by 
ALICE Collaboration \cite{Abelev:2013bla} and CMS Collaboration \cite{Chatrchyan:2012qb} 
as a function of $dn_{\rm ch}/dy$ (left panel) and a function of $Q_{\rm sat}$ (right panel).
Since ALICE and CMS select their events with different acceptance, different scaling functions appear.  
}\label{fig:7}
\end{figure}

Next, we turn our attention to the average transverse momentum $\langle p_{\rm T}\rangle$ obtained from the semi-inclusive distributions. 
The energy-dependent saturation momentum $Q_{\rm sat}(W^*)$ should be proportional to $\langle p_{\rm T} \rangle$ in GS framework \cite{Osada:2017oxe}. 
As seen in the left panel of Fig.\ref{fig:7},  
$dn_{\rm ch}/dy$ 
dependences of $\langle p_{\rm T} \rangle$ at 0.90, 2.76 and 7.00 TeV observed by 
ALICE \cite{Abelev:2013bla} and CMS \cite{Chatrchyan:2012qb} do not show scaling behavior in terms of $dn_{\rm ch}/dy$. 
However, since GS holds for the semi-inclusive distributions, one expects that $\langle p_{\rm T} \rangle$ is linearly proportional to 
$Q_{\rm sat}(W^*)$ and those data lie on a straight line regardless of the colliding energy $W$.  
Figure \ref{fig:7} shows results of the conversion of the $dn_{\rm ch}/dy$ dependence of $\langle p_{\rm T} \rangle$ on the left side panel 
into the dependence of $Q_{\rm sat}(W^*)$ on the right side panel. 
The difference in scaling curves between ALICE and CMS seems to be due to that in acceptance employed in each observation. 
Thus, the behavior of GS is observed not only in the inclusive distributions but also in the semi-inclusive distributions in high energy pp collisions. 
\subsection{Normalized fluctuation measure of transverse momentum}
A prominent scaling behavior emerges in event-by-event mean $p_{\rm T}$ fluctuations in pp collisions at LHC energies \cite{Abelev:2014ckr,Heckel:2015swa,Stefan:2011es}. 
In our previous work \cite{Osada:2017oxe}, we studied it focusing only on the energy $W=$ 0.90 TeV, 
and we did not discuss the GS behavior by extending the analysis to other energies. 
In this article, we analyze data on transverse momentum fluctuations observed at $\sqrt{s}=$ 0.90, 2.76, 7.00 TeV 
using $Q_{\rm sat}(W^*)$ and $S^*_{\rm T}$ without changing the basic idea of the model proposed in Ref.\cite{Osada:2017oxe}. 
The fluctuation measure is essentially a two-particle distribution as defined below, 
\begin{eqnarray}
  C_m &=& \frac{\int\! d^2{\bf p}_{\rm T_1}\int\! d^2{\bf p}_{\rm T_2}}{m(m-1)}~ \nonumber \\
  && \quad 
  \frac{d^4n_{\rm ch}}{d{\bf p}^2_{\rm T_1}  d{\bf p}^2_{\rm T_2}}~ 
  (p_{\rm T_1}-\langle p_{\rm T}\rangle)(p_{\rm T_2}-\langle p_{\rm T}\rangle), 
  \label{eq:Cm} 
\end{eqnarray}  
where $m=\frac{dn_{\rm ch}}{d\eta} \times |\Delta\eta|$ is the multiplicity in the pseudo-rapidity window $ |\Delta\eta|$.   
Since the universal function of GS is essentially one particle distribution, 
two-particle correlation function \cite{Gavin:2008ev} as shown below is required to obtain 
the two-particle distribution in Eq.(\ref{eq:Cm}); 
\begin{eqnarray}
  C({\bf p}_{\rm T_1},{\bf p}_{\rm T_2} ) \equiv 
  {\frac{d^4n_{\rm ch}}{d{\bf p}^2_{\rm T_1} d{\bf p}^2_{\rm T_2}}} \Bigg/
  \frac{d^2n_{\rm ch}}{d{\bf p}^2_{\rm T_1}} \frac{d^2n_{\rm ch}}{d{\bf p}^2_{\rm T_2}}.  
\end{eqnarray}
It is known that a gluon two-particle correlation function 
takes the following simple geometrical form in the CGC / Glasma framework \cite{Dumitru:2008wn,Lappi:2009xa,Lappi:2010cp}, 
\begin{eqnarray}
  C_{\rm GFT}({\bf p}_{\rm T_1},{\bf p}_{\rm T_2} )  =1+\frac{\kappa_2}{S_{\rm T}~Q_{\rm sat}^2}, 
  \label{eq:CGC/Glasma_flax_tube}
\end{eqnarray}
where $\kappa_2$ is a non-perturbative constant, and the evaluation of this constant is a challenging problem in theoretical physics. 
On the other hand, we consider an extreme model in which the correlation in momentum space between gluons 
is inherited to that between hadrons in the final state.
Since the transverse size of color flux tubes stretching between the receding protons is expected to be of order in $1/Q_{\rm sat}$, 
one may write the following correlation function commonly found in Bose-Einstein correlation (BEC) analysis: 
\begin{eqnarray}
  && C({\bf p}_{\rm T_1},{\bf p}_{\rm T_2} )  = \nonumber \\ 
  && \qquad 
  1+ \Big(S^*_{\rm T} ~[\kappa Q_{\rm sat}]^2\Big)^n
  \exp \left(-\frac{({\bf p}_{\rm T_1}-{\bf p}_{\rm T_2})^2}{\sigma [\kappa Q_{\rm sat}]^2}\right), \quad 
  \label{eq:BEC_correlation}
\end{eqnarray}
where $n$ and $\sigma$ are model parameters.   
Here, $\kappa$ is the parameter that appears in the universal function Eq.(\ref{eq:universal_func_fit}) 
which connects intermediate energy scale $Q_{\rm sat}$ and hadronization energy scale $T_{\rm eff}$. 
Since $\kappa$ and $Q_{\rm sat}$ always appear together in the inclusive distribution, 
there must also be such property in the two particle distribution in Eq.(\ref{eq:BEC_correlation}). 
Note also that the term $S^*_{\rm T} Q^2_{\rm sat}$ in Eq.(\ref{eq:BEC_correlation}) is proportional to the number of flux tubes \cite{Tribedy:2010ab},  
especially when $n = -1$, it can be interpreted as chaoticity of the BEC effect\cite{Osada:2017oxe}. 
Another parameter $\sigma$ is for adjusting the size of the flux tube. 
When $\sigma \approx 1$, it means that the size of the color flux tube is expanded by about $1/\kappa\approx 7.7$ 
times in the transverse direction and the source size scale is the inverse of the temperature of the system $\sim 1/T_{\rm eff}$. 
\begin{figure*}
\centerline{\includegraphics[width=13.0cm]{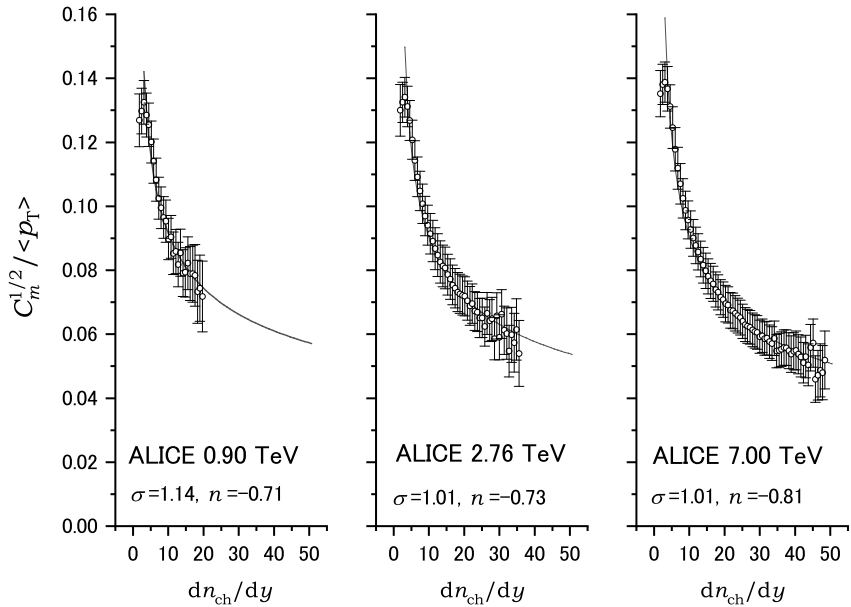}}
\caption{
Experimental data on event-by-event mean transverse momentum fluctuation \cite{Abelev:2014ckr} 
and fit results of our model (Eq.(\ref{eq:BEC_correlation})) to the data. 
The pseudo-rapidity window is $|\Delta\eta|=0.8$. 
Table \ref{tab:cmpt} shows the values ​​of the parameters that give the least chi-square for the fitting of $\sqrt{C_m}/\langle p_{\rm T}\rangle$.
We have excluded two  small $dN_{\rm ch}/dy$ (=1.8 and 2.4) data points from the fits.}\label{Fig:8}
\end{figure*}
As seen in Fig.\ref{Fig:8}, ALICE observed normalized fluctuation measure $\sqrt{C_m}/\langle p_{\rm T}\rangle$ at $W=$0.90, 2.76 and 7.00 TeV, 
and they found almost no energy dependence in them. 
Our model based on GS easily explain the reason why the measure $\sqrt{C_m}/\langle p_{\rm T}\rangle$ hardly depends on the collision energy: i.e., 
By noting that $p_{\rm T}=Q_{\rm sat} \tau^{1/(2+\lambda)}$, $\langle p_{\rm T}\rangle \propto Q_{\rm sat}$, and $m\propto S^*_{\rm T}Q_{\rm sat}^2$, 
one can represent the measure as a function of the scaling variable $\tau$ except for the term $S^*_{\rm T} Q_{\rm sat}^2$ in the correlation function Eq.(\ref{eq:BEC_correlation}). 
However, as shown by Eqs.(\ref{eq:Qsat_dndy}) and (\ref{eq:Rt_dndy}), the energy dependence of both $Q_{\rm sat}$ and $S^*_{\rm T}$ are considerably small. 
Moreover, recall that $S^*_{\rm T} Q_{\rm sat}^2$ is the number of color flux tubes. Since the gluon in the incident proton is saturated regardless of the energy, it is natural that the energy 
dependence of this factor is small. Therefore, it is explained that $\sqrt{C_m}/\langle p_{\rm T}\rangle$ is almost independently of the colliding energy $W$ in our model. 
The fit results to the experimental data by Eq.(\ref{eq:BEC_correlation}) are shown by solid lines in Fig.\ref{Fig:8}. 
We also show values of the parameter both $\sigma$ and $n$ giving $\chi^2_{\rm min}$ in Table \ref{tab:cmpt}. 
The values of $n$ obtained by the fits are from $-0.71$ to $-0.81$, which are larger than $-1$, 
but Eq.(\ref{eq:BEC_correlation}) can be compared with the Eq.(\ref{eq:CGC/Glasma_flax_tube}) in the Glasma framework. 
Evaluating the typical momentum scale of BEC as $|{\bf p}_{\rm T_1}-{\bf p}_{\rm T_2}|^2\sim [2\kappa Q_{\rm sat}]^2 \approx [200{\rm MeV}]^2$, 
the comparison 
leads us to a rough estimation of $\kappa_2$ as the following; 
\begin{eqnarray}
 \kappa_2 \sim \frac{1}{\kappa^2} \exp \left(-\frac{4}{\sigma}\right). 
\label{eq:k2_estimation}
\end{eqnarray}
Table \ref{tab:cmpt} also shows the values of $\kappa_2$ evaluated by Eq.(\ref{eq:k2_estimation}).   
Since there are considerable variations in the extracted values of $\kappa_2$ from experimental data based on the Glasma framework, 
its value is not known to be as accurate as an order of 1 \cite{Lappi:2009xa}. 
It is interesting to note that the values of $\kappa_2$ extracted from our model are comparable 
to the estimation by the Glasma framework, although the picture for particle correlation of each other is different.    
\begin{table}
\begin{ruledtabular}
\caption{\label{tab:cmpt}
The best fit values of the parameters $\sigma$ and $n$ to the experimental data 
on the event-by-event fluctuation of mean $p_{\rm T}$ observed by ALICE Collaboration \cite{Abelev:2014ckr} and 
results of evaluation of $\kappa_2$ by Eq.(\ref{eq:k2_estimation}).} 
\begin{tabular}{ccccl}
$\sqrt{s}$  [TeV] & $\sigma$ & $n$ & $\chi^2_{\rm min}$/dof & $\kappa_2$ \\ \hline 
0.90 & 1.14 & $-0.71$ & 4.93/23 & 1.79 \\ 
2.76 & 1.01 & $-0.73$ & 14.0/45 & 1.14 \\ 
7.00 & 1.01 & $-0.81$ & 20.4/64 & 1.14 \\ 
\end{tabular}
\end{ruledtabular}
\end{table}
\section{Summary and concluding remarks}\label{sec:4} 
In this article, we have phenomenologically investigated multiplicity dependence on the gluon saturation momentum in high energy pp collisions.  
This result makes it possible to classify events by energy-dependent saturation momentum $Q_{\rm sat}(W^*)$, 
which in turn can provide a new research approach to high energy multi-particle production. 

If the local parton-hadron duality hypothesis is correct,  
$Q_{\rm sat}(W^*)$ must link to observables in the final state of the charged hadrons. 
In order to extract $Q_{\rm sat}(W^*)$ that governs the multiplicity of the final states,
we assumed the semi-inclusive transverse momentum spectra exhibit geometrical scaling behavior independently of its fixed multiplicity and its colliding energy. 
Furthermore, the universal function is assumed to be the same as that of the inclusive distribution. 
Through the effective energy $W^*$ defined by Eq.(\ref{eq:W**}), 
we determined $Q_{\rm sat}(W^*)$ for the semi-inclusive distributions. 
We have shown that the transverse momentums distribution of various multiplicity class at $\sqrt{s}$=0.90, 2.76 and 7.00 TeV 
do scale in terms of the scaling variable $\tau^{1/(2+\lambda)}=p_{\rm T}/Q_{\rm sat}(W^*)$. 
We have also confirmed that  $Q_{\rm sat}(W^*)$ dependence on the average transverse momentum also scales 
to a linear function of $Q_{\rm sat}(W^*)$, which is consistent with the behavior expected from GS.
 
It is meaningful to note on works by Korus and Mr\'owczy\'nski \cite{Korus:2001fv,Mrowczynski:2004cg} and to compare with the model we have proposed. 
Korus and Mr\'owczy\'nski have introduced a multiplicity-dependent temperature 
and related the nontrivial behavior of fluctuations in the transverse momentum to that in the multiplicity distribution. 
In our model, on the other hand, the energy-dependent saturation momentum $Q_{\rm sat}(W^*)$ is related to the multiplicity of 
the final state via the effective energy $W^*$ and is also related to the temperature evaluated from the semi-inclusive spectra by Eq.(\ref{eq:T=kQ}). 
About fluctuation of transverse momentum, Korus and Mr\'owczy\'nski argue that the reason for it is 
that the fluctuation in the multiplicity distribution is almost independent of energy.
In fact, the normalized $q$-moment values of $C_2, C_3, C_4$ for the multiplicity distribution in the central rapidity region $|\eta|<0.5$ 
are almost independent of the collision energy \cite{Adam:2015gka,Khachatryan:2010nk}. 
On the other hand, in our model, the reason why there is almost no dependence on collision energy in the fluctuation measure 
$\sqrt{C_m}/\langle p_{\rm T}\rangle$ is that the energy dependences on $Q_{\rm sat}(W^*)$ and $S_{\rm T}$ are considerably 
small (see, Fig.\ref{fig:6}) in addition to the fact that the semi-inclusive transverse momentum spectrum shows the behavior of geometrical scaling. 

In this article, we thought that the two-particle Bose-Einstein correlation between identical gluons produced from 
color flux tubes could explain the experimental results of the fluctuation measure. 
The measure $\sqrt{C_m}/\langle p_{\rm T}\rangle$ can be fitted by Eq.(\ref{eq:BEC_correlation}) nicely, 
in which the correlation between gluons is considered to remain between charged particles after hadronization. 
Comparing Eq.(\ref{eq:CGC/Glasma_flax_tube}) with our model Eq.(\ref{eq:BEC_correlation}) 
we can estimate the value of the non-perturbative constant of the gluon correlation function $\kappa_2$. 
If a typical value for $|{\bf p}_{\rm T_1}-{\bf p}_{\rm T_2}|$ in Eq.(20) as 200 MeV/$c$ is adopted, one obtain $\kappa_2 = 1.4-1.8$.
It is interesting to extract $Q_{\rm sat}$ from other reaction such as pA \cite{McLerran:2015lta} and A-A\cite{Andres:2012ma} collision 
and to discuss the relationship between the fluctuation of multiplicity and that of the saturation momentum. 
However, we plan to investigate those issues at some other opportunity.\\

\begin{acknowledgements}
We acknowledge stimulating discussions with Grzegorz Wilk concerning 
the $q$-scaling in high-energy production processes and the effective energy. 
We would also like to thank the anonymous referee for thorough comments which have greatly improved the manuscript. 
\end{acknowledgements}
\bibliography{osada1967_2019}
\end{document}